\documentstyle[prb,twocolumn,aps]{revtex} 
\begin{document} 
\twocolumn[\hsize\textwidth\columnwidth\hsize\csname
@twocolumnfalse\endcsname

\title{Magnetic field -- temperature phase diagram of the organic 
conductor $\alpha $-(BEDT-TTF)$_2$KHg(SCN)$_4$} 
\author{P.~Christ$^1$, W.~Biberacher$^1$, M.V.~Kartsovnik$^{1,2}$, 
E.~Steep$^3$, E.~Balthes$^3$, H.~Weiss$^3$ and H.~M\"uller$^4$} 
\address{$^1$ Walther-Meissner-Institut, Walther-Meissner-Str. 8, D-85748 
Garching, Germany} 
\address{$^2$ Institute of Solid State Physics, 142432 Chernogolovka, 
Russia} 
\address{$^3$ High Magnetic Field Laboratory, MPI-FKF and CNRS,  
F-38042 Grenoble, France} 
\address{$^4$ European Synchrotron Radiation Facility, F-38043 Grenoble, 
France} 

\date{\today} 
\maketitle  

\begin{abstract} 
We present systematic magnetic torque studies of the ``magnetic field -- 
temperature'' phase diagram of the layered organic conductor 
$\alpha $-(BEDT-TTF)$_2$KHg(SCN)$_4$ at fields nearly perpendicular 
and nearly parallel to the highly conducting plane. The shape of the phase 
diagram is compared to that predicted for a charge-density-wave system 
in a broad field range. 
\end{abstract} 


\pacs{} ]\narrowtext
Organic metals $\alpha $-(BEDT-TTF)$_{2}M$Hg(SCN)$_{4}$, 
where $M=$ K, Tl, or Rb \cite{K-salt}, have attracted much 
attention in the last decade due to their exotic low-temperature electronic 
state. They are characterized by a layered crystal structure and a 
unique co-existence of quasi-one-dimensional (Q1D) and 
quasi-two-dimensional (Q2D) conducting bands \cite{K-salt}. 
The transition into the low-temperature state is associated with a 
$2k_{F}$ nesting instability of the Q1D part of the Fermi surface. 
Indeed, experiments on the angle-dependent magnetoresistance 
oscillations \cite{KKK,KAMRO,SasaAMRO,SingAMRO} 
have revealed a significant change in the electronic system due to a 
periodic potential with the wave vector close to the doubled Fermi wave 
vector of the Q1D band. On the other hand, studies of the magnetization 
anisotropy \cite{Sasakhi,Chrikhi}\ and $\mu $SR 
\cite{Pratt} give evidence for a low amplitude modulation of the 
magnetic moment suggestive of a spin-density wave (SDW). Many of 
the striking anomalies displayed by these compounds in magnetic field 
can be fairly well explained by the density-wave instability, taking into 
account the coexistence of the Q1D and Q2D Fermi surfaces (see e.g. 
Refs. \cite{KAMRO,SasaAMRO,SingAMRO,Mashov,McKenzie}). 
However, there remain several questions which can hardly be understood 
within the SDW model. One of  the important questions concerns the 
effect of magnetic field on the low-temperature state. 

It is known that magnetic field applied perpendicular to the direction 
of the spin polarization may stimulate the SDW formation in systems with 
imperfectly nested Fermi surfaces due to effective reduction of the 
electron motion to one dimension \cite{Gorkov,Montambaux}. This orbital 
effect leads to a slight increase of the SDW transition temperature as was 
shown for a Q1D conductor (TMTSF)$_{2}$PF$_{6}$ \cite{Danner}. The 
situation with the $\alpha $-(BEDT-TTF)$_{2}M$Hg(SCN)$_{4}$ salts is 
rather controversial in this respect. In agreement with the SDW model, 
Sasaki {\em et al.} \cite{SasaBT} reported the transition temperature, 
$T_{p}$, in $\alpha $-(BEDT-TTF)$_{2}$KHg(SCN)$_{4}$ to increase 
in magnetic field perpendicular to the spin polarization plane (which is the 
highly-conducting $ac$-plane in this compound \cite{Chriplane}). On the 
contrary, numerous other experiments suggest a reduction of $T_{p}$ in 
magnetic field. Some authors \cite{House,Brooks} claim that the 
low-temperature state is completely suppressed in this salt and the normal 
metallic state is restored above the so-called {\em kink} transition at 
$B_{\text{kink}}\simeq 24$~T. On the other hand, several works suggest 
a new phase, different from the normal one, to emerge above 
$B_{\text{kink}}$ \cite{KartBT,Biskup}. Based on the shape of the 
``magnetic field -- temperature'' ($B-T$) phase diagrams 
\cite{KartBT,Biskup}, Biskup {\em et al.} \cite{Biskup} proposed the 
phase transition to be driven by a charge-density wave (CDW) rather than 
SDW instability. 

It should be noted, that the studies of the high-field region of the 
$B-T$ diagram of the $\alpha $-(BEDT-TTF)$_{2}M$Hg(SCN)$_{4}$ 
compounds have been mostly done by use of magnetoresistance technique. 
Obviously, such experiments are difficult to interpret unambigously in terms 
of phase transitions. Therefore a detailed investigation of thermodynamic 
properties is necessary in order to establish the phase boundaries. So far 
only few magnetization data at fields above 15~T were presented in two 
works \cite{SasaBT,KartBT}. However, the conclusions made in these works 
concerning the field effect on the transition temperature contradict each 
other. To elucidate the problem, we have carried out a systematic study of 
the $B-T$ phase diagram of $\alpha $-(BEDT-TTF)$_{2}$KHg(SCN)$_{4}$ by 
means of magnetic torque experiments. 

Several high quality samples chosen for the experiment were grown by the 
standard electrochemical method \cite{Chemistry} and had a typical mass of 
100 to 350 $\mu $g. A cantilever beam magnetometer \cite{Chrikhi} was used 
to measure the torque in fields nearly perpendicular and nearly 
parallel to the highly conducting $ac$-plane. The measurements were 
performed at temperatures between 0.4 and 18~K in magnetic 
fields up to 28~T produced at the High Magnetic Field Laboratory 
in Grenoble, France. 

We first focus on field directions almost perpendicular to the layers. 
Typical field dependencies of the steady part of the torque 
$\tau _{\text{st}}(B)$ 
are shown in Fig. 1a, for the angle $\theta $ between the magnetic 
field and the normal to the $ac$-plane equal to 2.2$^{\circ }$. At high 
temperatures ($T\geq $ 8~K) we find an almost temperature insensitive 
quadratic dependence of the torque on magnetic field. On lowering the 
temperature below 8~K the quadratic term increases at small fields, but 
above 4~T the dependence becomes weaker than quadratic and at 
high fields the curves bend to merge with the high temperature curve. The 
field at which the torque returns back to its normal behaviour coincides 
with the kink field  $B_{\text{kink}}$ as determined in other experiments 
\cite{SasaBT,House,Brooks,KartBT,Biskup}. In addition to the steady part of 
the torque, de Haas-van Alphen (dHvA) oscillations were observed. At 10~K 
these oscillations were resolved only at the highest fields, but at 5.0~K 
their amplitude was already comparable to $\tau _{\text{st}}(B)$ as shown 
by a dashed line in Fig.~1a. To extract $\tau _{\text{st}}(B)$ 
we used a Fourier filter. In contrast to the measurements at 
higher temperatures, the curve at 3.2~K does not return to the high 
temperature part at $B_{\text kink}$ 
but stays below. For temperatures below 3~K the dHvA 
amplitude becomes so strong that the steady torque cannot be extracted 
reliably any more. In Fig.~1b we show a trace of a field sweep from 18~T 
to 28~T and back made at 0.4~K. There is a clear transition from a 
low field state (characterized by a splitting of the oscillation amplitude) 
to a high field state (characterized by a higher oscillation amplitude and 
the absence of splitting). This transition shows a strong hysteresis of the 
dHvA amplitude in the field interval marked by fat arrows in Fig.~1b. 
Furthermore, there is a significant shift between up and 
down sweep curves in the high field part indicating a complex magnetic 
state. 

To clarify the latter point, we performed temperature sweeps at 
constant fields. For these experiments it is of crucial importance to 
suppress the influence of the oscillatory part \cite{KartBT}. We therefore 
performed these sweeps at field values, at which the dHvA contribution to 
the temperature dependence is nearly zero. The results are shown in 
Fig.~2a. Despite a small remanent dHvA contribution there is still a 
clear transition into a new state even at the highest field. 

In order to determine anisotropy effects in the phase diagram, we performed 
torque experiments at fields almost parallel to the layer plane. The 
phase transition is clearly seen in temperature sweeps. Typical examples 
taken at different fields at $\theta =$~87.5$^{\circ }$ are given in 
Fig.~2b. The field dependence of the torque below 4 K shows a complex 
behavior with a strong hysteresis between up and down field sweeps 
\cite{pchrist}. This behavior is drastically different from the feature 
observed at the kink transition at low angles.  An example of a field 
sweep at 1.3~K is shown in the inset in Fig.~2b.

The results of our studies can be summarized by plotting a $B-T$ phase 
diagram as shown in Fig.~3. Here the data obtained on 4 samples having 
slightly different $T_{p}$ (ranging from 8.0 to 8.4~K) are presented. 
That is why the temperature and field are given in reduced units $T/T_{p}(0)$ 
and $\mu _{B}B/k_{B}T_{p}(0)$, respectively [here $T_{p}(0)$ is the 
extrapolated critical temperature at zero field]. 
The definition of the transition points is illustrated in Figures 1 and 2.

The low-angle data in Fig.~3 are qualitatively consistent with the $B-T$ 
diagrams obtained from earlier magnetoresistance \cite{KartBT,Biskup} and 
torque \cite{KartBT} measurements in tilted fields: Firstly, the transition 
temperature continuously decreases with increasing the field; secondly, the 
low-temperature state is different from the normal non-magnetic state even 
above the kink transition. Quantitatively, our data are in perfect agreement 
with those obtained from specific heat measurements at $B\leq 14$~T 
\cite{Kovalev}. These results are obviously in conflict with the SDW model. 
On the other hand, they can be compared to what is expected for a CDW 
\cite{Zanchi}. At low field the CDW$_{0}$ phase with an optimal zero-field 
wave vector is stable below $T_{p}$. As the field increases, the Zeeman 
splitting of the subbands with antiparallel spins leads to the deterioration of 
the nesting conditions and, consequently, suppression of $T_{p}$ 
\cite{Pauli}. However, when the Zeeman splitting energy reaches the value of 
the zero-temperature energy gap, a formation of a spatially modulated 
CDW$_{x}$ state with a longitudinally shifted wave vector is expected. This 
state is analogous to the Fulde-Ferrel-Larkin-Ovchinnikov state predicted 
for superconductors \cite{FFLO} and persists to considerably higher fields 
than the conventional CDW$_{0}$. The phase diagram proposed by Zanchi {\em 
et al.} \cite{Zanchi} for a CDW system with perfect nesting is shown by 
dashed lines in Fig.~3. Apart from different field scales, the phase 
diagrams are remarkably similar to each other. 

Assuming the CDW model, the deviation of the actual phase boundary 
for fields nearly perpendicular to the plane to higher temperatures at 
$T_p/T_p(0)>0.6$ can be ascribed to a significant orbital effect of the 
magnetic field. This effect is important for an imperfectly nested 
Fermi surface and leads to a relative increase of $T_{p}$ 
\cite{Montambaux,Zanchi}. In our case, when the warping of the open Fermi 
surface sheets is much stronger within the $ac$-plane than in the interlayer 
direction, the orbital effect should be anisotropic: its contribution 
decreases as the angle $\theta $ approaches $90^{\circ }$. Indeed, the 
critical temperature of the transition into the low-temperature low-field 
state is found to be systematically lower at $\theta \simeq 90^{\circ }$, 
lying perfectly on the theoretical line (Fig.~3). This implies that the 
orbital effect is absent for the in-plane field direction. 

In the high-field region, the phase lines determined at different field 
orientations seem to converge, suggesting an isotropic effect of magnetic 
field on the transition temperature into the low-temperature high-field 
state. For a definite conclusion, more detailed studies at different angles 
are needed. 

The considerable difference between the field scale in the phase diagram 
obtained from the experiment and that predicted by the CDW model 
is not very surprising. Indeed, the model calculations 
\cite{Zanchi} are made within a mean-field approximation neglecting 
fluctuation effects. The latter may significantly lower $T_{p}(0)$ with 
respect to the mean-field value. Furthermore, the imperfect nesting which 
likely occurs in the present system has a stronger suppressing effect on 
$T_{p}(0)$ than on the critical field \cite{Zanchi}. Both these factors lead 
to an underestimation of the actual critical fields. 

Finally, we note that the field dependence of the torque at high angles 
has no simple explanation 
within the proposed model. The non-monotonic torque with a 
hysteresis between up and down field sweeps observed at $\theta \gtrsim 
60^{\circ }$ \cite{pchrist} is reminiscent of multiple phase transitions. As 
the angle approaches 90$^{\circ }$, the features become less 
pronounced though still persist to the angles as high as 88-89$^{\circ }$ 
(see inset in Fig.~2b). In principle, an additional phase transition into a 
CDW$_{y}$ state with a transversally shifted wave vector may be expected at 
high angles at which the orbital effect is sufficiently suppressed 
\cite{Zanchi}. Still, it cannot account for the whole structure of the 
torque at high angles and its complicated angular dependence. Obviously, the 
applied model \cite{Zanchi} is too oversimplified to explain all the field 
effects. For a more adequate description it seems very important to include 
the Q2D band into consideration. In particular, it was recently shown 
that oscillations of the chemical potential due to the quantization of the 
2D orbits have a significant impact on the CDW gap \cite{Harrison}. 
On the other hand, the magnetization anisotropy itself, 
revealing an ``easy-plane'' spin polarization at low temperatures 
\cite{Chriplane}, indicates a non-trivial magnetic structure linked to the 
probable CDW. 

In conclusion, we have presented a $B-T$ phase diagram of 
$\alpha $-(BEDT-TTF)$_{2}$KHg(SCN)$_{4}$ 
built on the basis of magnetization measurements. The shape of the diagram 
and the effect of the field orientation are suggestive of a CDW formation 
accompanied by imperfect nesting of the Q1D part of the Fermi surface. If 
this is true, the high-field phase would represent the first example of a 
CDW with a spatially modulated wave vector.

We thank A. Bjeli\v{s} for very useful discussions. The work was 
supported in part by the TMR Program of the European Community, contract 
No. ERBFMGECT950077.

\mbox{}

{\bf Figure captions}

Fig. 1. Torque as a function of  magnetic field applied nearly 
perpendicular to the $ac$-plane: (a) - steady part of the torque at 
different temperatures; the dotted curve represents the total signal from 
the sample, with the dHvA oscillations at $T=5.0$~K; 
(b) - up (dotted line) and down (solid line) field sweeps of the torque 
at low temperature. 

Fig. 2. Temperature sweeps of the torque at $\theta = 2.2^{\circ }$ (a) 
and $87.5^{\circ }$ (b) at different fields. The inset shows the field 
dependence of the torque at $\theta = 87.5^{\circ }$, $T=1.3$~K. 

Fig. 3. Phase diagram of $\alpha $-(BEDT-TTF)$_2$KHg(SCN)$_4$. 
Different symbols correspond to the transition points obtained from: 
the $\tau_{\text st}(T)$ sweeps at $\theta =2.2^{\circ }$ (stars, sample 
\#1), 6.5$^{\circ }$ (solid diamonds, sample \#2), 11.8$^{\circ }$ (solid 
up-triangles, sample \#3), 87.5$^{\circ }$ (open squares, sample 
\#1), and 89.5$^{\circ }$ (open up triangles, sample \#1); 
$\tau_{\text st}(B)$ sweeps at $\theta =2.2^{\circ }$ (crosses, sample 
\#1); and characteristic changes in the dHvA signal at 
$\theta =4.0^{\circ }$ (solid down-triangles, sample \#4). 
The dashed lines represent the phase diagram predicted for a CDW 
system with a perfectly nested Fermi surface \cite{Zanchi}.


\begin{references}
\bibitem{K-salt} H.~Mori, S.~Tanaka, M.~Oshima et al., Bull. 
Chem. Soc. Jpn. {\bf 63}, 2183 (1990). 

\bibitem{KKK} M. V.~Kartsovnik, A. E.~Kovalev and N. D.~Kushch, 
J. Phys I France {\bf 3}, 1187 (1993).

\bibitem{KAMRO} A.E.~Kovalev, M.V.~Kartsovnik, R.P.~Shibaeva 
et al., Solid State Commun. {\bf 89}, 575 (1994).


\bibitem{SasaAMRO} T.~Sasaki and N.~Toyota, Phys. Rev. B {\bf 49}, 
10120 (1994). 

\bibitem{SingAMRO} J.~Caulfield, S.J.~Blundell, M.S.L.~du Croo 
de Jongh et al., Phys.~Rev.~B {\bf 51}, 8325 (1995). 

\bibitem{Sasakhi} T.~Sasaki, H.~Sato, and N.~Toyota, Synth. Met. 
{\bf 41-43}, 2211 (1991). 

\bibitem{Chrikhi} P.~Christ, W.~Biberacher, H.~M\"uller, and 
K.~Andres, Solid State Commun. {\bf 91}, 451 (1994). 

\bibitem{Pratt} F.L.~Pratt, T.~Sasaki, N.~Toyota, and K.~Nagamine, 
Phys. Rev. Lett. {\bf 74}, 3892 (1995). 

\bibitem{Mashov} M.V.~Kartsovnik, D.V.~Mashovets, D.V.~Smirnov 
et al., J. Phys. I France {\bf 4}, 159 (1994). 

\bibitem{McKenzie} R.H.~McKenzie, G.J.~Athas, J.S.~Brooks 
et al., Phys. Rev. B \textbf{54}, R8289 (1996). 

\bibitem{Gorkov} L.P.~Gor'kov and A.G.~Lebed, J. Phys. Lett. 
(Paris) {\bf 45}, L433 (1984). 

\bibitem{Montambaux}  G.~Montambaux, Phys. Rev. B {\bf 38}, 4788 
(1988). 

\bibitem{Danner}  G.M. Danner, P.M.~Chaikin, and S.T.~Hannahs, Phys. Rev. B 
{\bf 53}, 2727 (1996). 

\bibitem{SasaBT} T.~Sasaki, A.G.~Lebed, T.~Fukase, and N.~Toyota, 
Phys. Rev. B {\bf 54}, 12969 (1996). 

\bibitem{Chriplane}  P. Christ, W. Biberacher, W. Bensch et al., 
Synth. Met. {\bf 86}, 2057 (1997). 

\bibitem{House}  A.A.~House, S.J.~Blundell, M.M.~Honold 
et al., J.~Phys. Condens. Matter {\bf 8}, 8829 (1996). 

\bibitem{Brooks} J.S.~Brooks, X.~Chen, S.J.~Klepper et al., 
Phys. Rev. B {\bf 52}, 14457 (1995). 

\bibitem{KartBT} M.V.~Kartsovnik, W.~Biberacher, E.~Steep et al., 
Synth. Met. {\bf 86}, 1933 (1997). 

\bibitem{Biskup}  N.~Biskup, J.A.A.J. Perenboom, J.S.~Brooks, 
and J.S.~Qualls, Solid State Commun. {\bf 107}, 503 (1998). 

\bibitem{Chemistry} H.~M\"uller and Y.~Ueba, Synthesis {\bf 9}, 
853 (1993). 

\bibitem{pchrist} P.~Christ, W.~Biberacher, A.G.M.~Jansen et al.,
 Synth. Met. {\bf 70}, 823 (1995). 

\bibitem{Kovalev}  A.E.~Kovalev, H.~M\"{u}ller, and M.V.~Kartsovnik, 
Zh. Eksp. Teor. Fiz. {\bf 113}, 1058 (1998) [JETP {\bf 86}, 578 (1998)]. 

\bibitem{Zanchi} D.~Zanchi, A.~Bjeli\v{s} and G.~Montambaux, 
Phys. Rev. B \textbf{53}, 1240 (1996). 

\bibitem{Pauli} W.~Dietrich and P.~Fulde, Z. Phys. {\bf 265}, 239 (1973). 

\bibitem{FFLO} P.~Fulde and R.A.~Ferrel, Phys. Rev. {\bf 135}, A550 (1964); 
A.I.~Larkin and Yu.N.~Ovchinnikov, Sov. Phys. JETP {\bf 20}, 762 (1965). 

\bibitem{Harrison}  N.~Harrison, Phys. Rev. Lett. \textbf{83}, 1395 (1999). 
\end{references}
\end{document}